\documentclass[prl,twocolumn,showkeys,amsmath,amssymb]{revtex4}

\usepackage{graphicx}

\newcommand{\vecr}{\mathbf{r}}

\begin{document}

\title{Algebraic perturbation theory for dense liquids with discrete potentials}

\author{Artur B. Adib}
\email{adiba@mail.nih.gov}
\affiliation{
Laboratory of Chemical Physics, NIDDK, National Institutes of Health, Bethesda, Maryland 20892-0520, USA
}

\date{\today}

\begin{abstract}
A simple theory for the leading-order correction $g_1(r)$ to the structure of a hard-sphere liquid with discrete (e.g. square-well) potential perturbations is proposed. The theory makes use of a general approximation that effectively eliminates four-particle correlations from $g_1(r)$ with good accuracy at high densities. For the particular case of discrete perturbations, the remaining three-particle correlations can be modeled with a simple volume-exclusion argument, resulting in an {\em algebraic} and surprisingly accurate expression for $g_1(r)$. The structure of a discrete ``core-softened'' model for liquids with anomalous thermodynamic properties is reproduced as an application.
\end{abstract}

\keywords{High-temperature expansion; Solvation theory; Solute-solvent interaction}

\maketitle


Although in principle the properties of complex physical systems can be computed to arbitrary accuracy through computer simulations, perturbation theory remains instrumental across quantum, classical and statistical mechanics. Its utility is apparent whenever one is interested in getting quantitative insights into the behavior of a system under different conditions, a task that would otherwise require several -- possibly costly -- simulations spanning the desired parameters. The underlying assumption is that there exists a well-characterized reference system, the properties of which are relatively similar to that of the system of interest, so that discrepancies between the two can be quantified through a small parameter expansion. The use of a perturbative approach is then justified, as is often the case, if the leading-order terms in this expansion can be computed with greater ease than those required by the exact approach.

When applied to the structure of simple atomic liquids, in which the interaction potential is comprised of a small, smooth and short-ranged attractive component outside a sharply repulsive core, liquid-state perturbation theory \cite{zwanzig54} takes on a especially simple form: According to the Weeks-Chandler-Andersen theory (see e.g. \cite{hansen86}), the structure of such systems is largely dominated by the repulsive part of the potential, and particularly at high densities it can be approximated by the structure $g_0(r)$ of the purely repulsive fluid alone with excellent accuracy \cite{chandler70}; in this rather peculiar case, perturbative corrections to the reference structure are hardly necessary \cite{weeks71,chandler83}.

For more complex liquid models with radial interactions other than such specific attractions, however, one cannot expect to always find simple reference systems with this property. Interest in such models stems in particular from recent efforts to model \cite{headgordon94,headgordon95,ashbaugh01} and explain \cite{stanley98prl,stanley01nature,stanley05prl,barbosa06a,barbosa06b} the properties of anomalous liquids through the existence of multiple length scales in the interaction potential, which introduce additional repulsive and attractive interactions outside the atomic core. In these and similar cases, perturbation theory needs to account for terms beyond the reference (repulsive-core) structure $g_0(r)$.

Unfortunately, the leading-order correction $g_1(r)$ to the radial distribution function of a liquid due to an arbitrary perturbing potential is in general a function of two-, three-, and four-particle correlations of the reference fluid (cf. \cite{hansen86,barker76}). Consequently, over the years several approximation schemes have been proposed to simplify this correction \cite{barker76,smith76,henderson76,tang94a,tang94b,largo05}. Such efforts have focused either on the superposition approximation, which eliminates high-order correlation functions but is known to fail at moderately high densities \cite{barker76}, or on the application of the integral equation method with different approximate closures \cite{barker76,smith76,henderson76,tang94a,tang94b,largo05}, which typically requires the numerical solution of a nonlinear integral equation for $g_1(r)$ at each condition of interest. In the particular case of square well potentials, the latter has been applied with varying degrees of success, depending on the density and the range of the well \cite{barker76,henderson76,largo05}.

The present paper introduces simple approximations that result in algebraic corrections to $g_0(r)$ (cf. Eq.~(\ref{g1-M})) when the perturbing potential is a discrete function of $r$ -- i.e. comprised of ``steps'' -- thereby avoiding the numerical computations required by previous methods while achieving competitive accuracies at high densities. The theory is illustrated for liquids with one \cite{barker76} and two \cite{stanley98prl,stanley01nature} steps in the interaction potential.

To begin the derivation, let $\lambda U(\vecr^N)$ be the $N$-body perturbation potential of a homogeneous, single-component fluid. The two-particle density can then be expanded in powers of $\lambda$, and a convenient expression for the leading-order correction $\lambda \rho^{(2)}_1(\vecr_1, \vecr_2)$ to the reference density $\rho_0^{(2)}(\vecr_1, \vecr_2)$ can be found by directly $\lambda$-differentiating the definition of the two-particle density in the canonical ensemble, yielding \cite{lowry64}
\begin{subequations} \label{exact}
\begin{equation} \label{rho-exact-U}
  \rho^{(2)}_1(\vecr_1, \vecr_2) = - \beta \rho_0^{(2)}(\vecr_1, \vecr_2) \left[ \langle U \rangle_0^{(1,2)} - \langle U \rangle_0 \right].
\end{equation}
Here $\beta = 1/kT$ is the usual temperature parameter, $\langle \cdot \rangle_0$ is the ensemble average with respect to the reference system, and $\langle \cdot \rangle_0^{(1,2)}$ is the same average with the constraint that particles 1 and 2 are held fixed at $\vecr_1$ and $\vecr_2$, respectively. As it stands, it can be straightforwardly shown that this expression is fully equivalent to the ``discrete representation'' of Barker and Henderson \cite{barker76} in the limit of infinitely thin shells, or to the more traditional result in terms of many-particle densities \cite{barker76,hansen86}; of course, use of either expression in analytic form presents a formidable challenge, as very little is known about the required many-particle correlation functions.

Let us now decompose $U$ in terms of one-particle energies $U(\vecr^N) = \frac{1}{2} \sum_{i=1}^N u_i(\vecr^N)$, where $u_i(\vecr^N) = \sum_{j\neq i} u(\vecr_i, \vecr_j)$ is the perturbation energy of particle $i$ with the remaining ones, and $u(\vecr_i, \vecr_j) = u(|\vecr_i-\vecr_j|)$ is the (spherically symmetric) pairwise perturbation energy between particles $i$ and $j$; Eq.~(\ref{rho-exact-U}) can thus be written as
\begin{multline} \label{rho-exact-u}
  \rho^{(2)}_1(\vecr_1, \vecr_2) = - \beta \rho_0^{(2)}(\vecr_1, \vecr_2) \left[  \langle u_1 \rangle_0^{(1,2)} - \langle u_1 \rangle_0 \right. \\
  \left. + \frac{N-2}{2} \left( \langle u_3 \rangle_0^{(1,2)} - \langle u_3 \rangle_0 \right) \right],
\end{multline}
\end{subequations}
where use was made of the indistinguishability of the particles wherever convenient. Consider now the last term in parentheses. Since the inhomogeneities caused by particles 1 and 2 occupy volumes of molecular size only, one expects the difference $\langle u_3 \rangle_0^{(1,2)} - \langle u_3 \rangle_0$ to scale down extensively with the total volume of the fluid. The first approximation to be invoked in this paper consists in assuming that the resulting (in principle intensive) contribution to Eq.~(\ref{rho-exact-u}) due to the average of $u_3$ is vanishingly small in comparison to that due to $u_1$, so that
\begin{subequations} \label{myapprox}
\begin{equation} \label{rho-approx-u}
  \rho^{(2)}_1(\vecr_1, \vecr_2) \approx - \beta \rho_0^{(2)}(\vecr_1, \vecr_2) \left[ \langle u_1 \rangle_0^{(1,2)} - \langle u_1 \rangle_0 \right].
\end{equation}
This approximation effectively eliminates four-body correlations from Eq.~(\ref{rho-exact-u}), and thus can be expressed in terms of pair and triplet particle densities only:
\begin{multline}
  \rho^{(2)}_1(\vecr_1, \vecr_2) \approx - \beta \rho_0^{(2)}(\vecr_1, \vecr_2) \left\{ u(\vecr_1, \vecr_2) \phantom{\frac{1}{1}} \right. \\
  \left. + \int \! \! d\vecr_3 \, u(\vecr_1, \vecr_3) \left[ \frac{\rho^{(3)}_0(\vecr_1, \vecr_2, \vecr_3)}{\rho_0^{(2)}(\vecr_1, \vecr_2)} - \frac{ \rho^{(2)}_0(\vecr_1, \vecr_3) }{\rho} \right] \right \},
\end{multline}
or, in terms of distribution functions,
\begin{equation}
  g_1(r) \approx - \beta g_0(r) \left[ u(r) + \rho \int \! \! d\vecr' \, u(r') \, \Delta g_0(\vecr' | \vecr) \right],
\end{equation}
\end{subequations}
where $\rho = N/V$ is the particle density, and $\Delta g_0(\vecr' | \vecr) \equiv g_0(\vecr' | \vecr ) - g_0( \vecr' )$ is the change in the local structure at~$\vecr'$ from a given fluid particle due to the presence of another particle fixed at~$\vecr$ from it. Note that Eq.~(\ref{myapprox}) would be exact had the perturbation potential acted on the total energy of particles 1 and 2 only, i.e. $\lambda U = \lambda (u_1+u_2)/2$, as for example in the theory of hydrophobicity, where particles 1 and 2 would play the role of solutes, and one would like to compute corrections due to solute-solvent and solute-solute interactions to solute-solute potentials of mean force \cite{pratt80}. With this analogy in mind, the approximation in Eq.~(\ref{myapprox}) is essentially equivalent to neglecting structural rearrangements due to ``solvent-solvent'' perturbations (i.e. interactions between particles $3,4,\ldots,N$); consequently, it is expected to be more accurate at high densities, wherein the relative arrangement of such particles is primarily determined by hard-sphere packing alone. These observations motivate referring to Eq.~(\ref{myapprox}) as the {\em solvation approximation} to $g_1(r)$.

\begin{figure}
\begin{center}
\includegraphics[width=240pt]{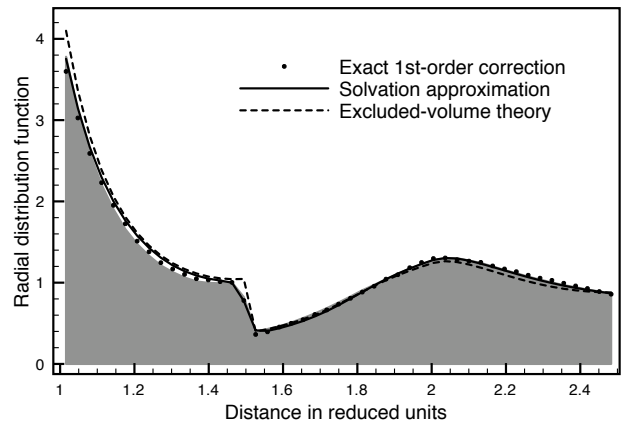} 
\caption{Radial distribution function for a hard-sphere fluid with a single square-well of depth $\varepsilon/kT=1$ extending from $r/\sigma=1$ to $r/\sigma = 1.5$, corrected according to the various expressions for $g_1(r)$ in the paper. The particle density is $\rho \sigma^3 = 0.85$. The shaded curve is the ``exact'' $g(r)$ obtained by direct Monte Carlo simulation. The exact 1st-order correction is computed with Eq.~(\ref{rho-exact-U}), the solvation approximation with Eq.~(\ref{rho-approx-u}), and the excluded-volume theory with Eqs.~(\ref{g1-exclvol})-(\ref{SW-deltan}).
}
\label{fig:results-SW}
\end{center}
\end{figure}

\begin{figure}
\begin{center}
\includegraphics[width=240pt]{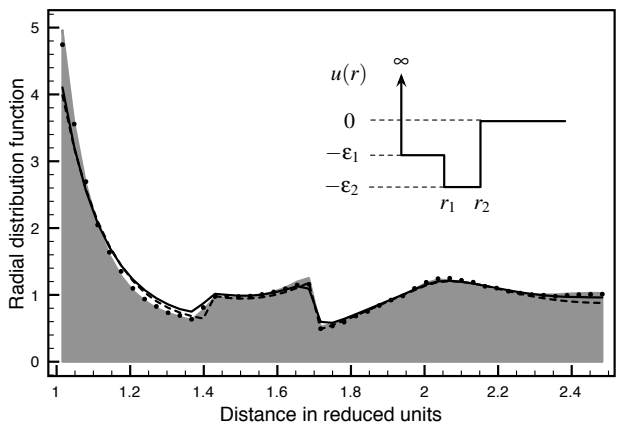} 
\caption{Radial distribution function for the double square-well potential of Ref.~\cite{stanley98prl} at $\rho \sigma^3 = 0.85$. Legends as in Figure~\ref{fig:results-SW}. The step depths are $\varepsilon_1/kT = 0.5$ and $\varepsilon_2/kT = 1$, corresponding to the steps ending at $r_1/\sigma = 1.4$ and $r_2/\sigma = 1.7$, respectively (inset). The excluded-volume theory is given by Eq.~(\ref{g1-M}) with $M=2$.
}
\label{fig:results-CS}
\end{center}
\end{figure}

\begin{figure}
\begin{center}
\includegraphics[width=240pt]{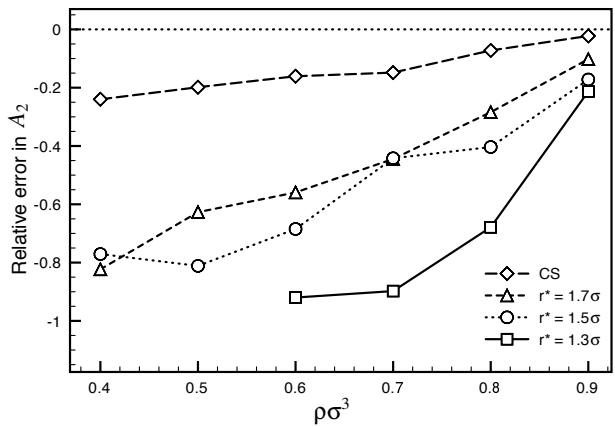} 
\caption{Relative error $(A_2^{solv} - A_2^{exact})/|A_2^{exact}|$ in the second-order correction to the free energy, viz. $A_2/N = (\rho / 4) \int d\vecr \, u(r) \, g_1(r)$ \cite{barker76}, versus density, where the superscripts specify the use of Eq.~(\ref{rho-exact-U}) [$A_2^{exact}$] or Eq.~(\ref{rho-approx-u}) [$A_2^{solv}$] for the computation of $g_1(r)$. From top to bottom: Discrete core-softened (CS) model of Figure~\ref{fig:results-CS}, and single-well potentials with ranges $r^*/\sigma=1.7$, $1.5$, and $1.3$.
}
\label{fig:A2}
\end{center}
\end{figure}

The validity of Eq.~(\ref{myapprox}) has been tested for both discrete and continuous potentials typically used in liquid-state perturbation theory -- such as the square-well liquid \cite{barker76}, and the Barker-Henderson and Weeks-Chandler-Andersen partitioning schemes for Lennard-Jones potentials \cite{hansen86} -- as well as for the discrete ``core-softened'' model of Ref.~\cite{stanley98prl}. As illustrated in Figs.~\ref{fig:results-SW} and \ref{fig:results-CS} for the discrete models of present interest, the solvation approximation provides good quantitative results for both one- and two-step potentials at high densities ($\rho \sigma^3 = 0.85$); in agreement with the above discussion, however, the quality of the approximation decreases at lower densities (cf. Fig.~\ref{fig:A2} and discussion on thermodynamics below). Although not central to the algebraic development below, it is also worth mentioning that the sampling convergence of Eq.~(\ref{rho-approx-u}) seems to be much faster than that of the exact expression, Eq.~(\ref{rho-exact-U}); it would be interesting to explore such issues in the context of the unbiased estimators of Ref.~\cite{adib01}.


For the particular case of discrete potentials, the solvation approximation is considerably simplified. Consider for example the square-well perturbation to a hard-sphere fluid where $u(r) = -\varepsilon$ for $r<r^*$, and $u(r)=0$ otherwise, with $r^*>\sigma$. Inserting this expression in Eq.~(\ref{myapprox}) yields
\begin{equation} \label{g1-exclvol}
  g_1(r) = -\beta g_0(r) \left[ u(r) - \varepsilon \,  \Delta n_0^*(r) \right],
\end{equation}
where $\Delta n_0^*(r)$ is the net change in the average number of particles inside a shell of radius $r^*$ surrounding a given particle, due to the presence of another particle fixed at a distance $r$ from it. The problem of computing the leading-order correction to $g_0(r)$ for such systems thus reduces to that of quantifying population changes in the coordination shell defined by the radius $r^*$ due to an ``intruder'' particle at $r$ in the reference fluid. In the following, a simple volume-exclusion theory for $\Delta n_0^*(r)$ will be presented.

\begin{figure}
\begin{center}
\includegraphics[width=120pt]{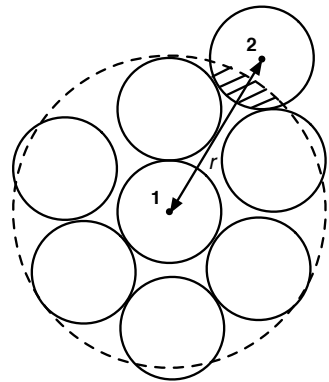} 
\caption{Two-dimensional representation of the excluded-volume model [Eq.~(\ref{SW-deltan})] for the single square-well potential with $r^* = 1.5 \sigma$. Solid circles represent the hard spheres of the reference system with diameter $\sigma$, while the larger dashed circle represents the outer radius $r^*$ of the perturbation well. The present model assumes that in the reference hard-sphere system, the average number of particles expelled from the $r^*$-shell of particle 1 due to the presence of particle 2 is proportional to the volume $v^*(r)$ of particle 2 inside the shell (shaded area). The normalization is such that exactly one particle is expelled from the $r^*$-shell when particle 2 is fully inside the well region delimited by the dashed circle.}
\label{fig:spheres}
\end{center}
\end{figure}

For the sake of clarity, let us focus on the single square-well case with $r^*=1.5 \sigma$, and let us monitor the average number of particles within the $r^*$-shell of particle 1 as particle 2 is brought to a distance $r$ from it (see Fig.~\ref{fig:spheres}). To a first approximation, appreciable changes in the average number of particles inside the $r^*$-shell of particle 1 should occur whenever particle 2 occupies a volume that would otherwise be occupied by the particles inside that shell. At sufficiently high densities, wherein the particles are tightly packed around each other, this should happen progressively as the body of particle 2 infiltrates the $r^*$-shell; a simple model that captures this idea is thus
\begin{equation} \label{SW-deltan}
  \Delta n_0^* (r) = - \frac{v^*(r)}{v_\sigma},
\end{equation}
where $v^*(r)$ is the two-sphere intersection volume shown in Fig.~\ref{fig:spheres}, and $v_\sigma = \pi \sigma^3 / 6$ is the volume of a sphere of diameter $\sigma$. Note that this expression is normalized so that $\Delta n_0(\sigma) = -1$, i.e. exactly one particle is excluded from the $r^*$-shell when particle 2 is fully inside the $r<r^*$ region. Since a polynomial expression for the intersection volume $v^*(r)$ of the spheres can be obtained by simple quadrature, Eqs.~(\ref{g1-exclvol})-(\ref{SW-deltan}) yield the desired algebraic theory for the leading-order correction to $g_0(r)$ in the particular single square-well case. Notwithstanding its simplicity, the resulting theory is in striking quantitative agreement with ``exact'' Monte Carlo results (Fig.~\ref{fig:results-SW}).

The above model can be directly extended to encompass multiple steps of arbitrary widths; thus, for a potential with $M$ steps,
\begin{equation} \label{g1-M}
  g_1(r) = -\beta g_0(r) \left[ u(r) + \sum_{i=1}^M \varepsilon_i \, \frac{v^{(i)}(r)}{v_\sigma} \right],
\end{equation}
where $-\varepsilon_i$ is the energy of the $i$th step, and $v^{(i)}(r)$ is the volume of the intruder particle inside the $i$th step region (defined analogously to $v^*(r)$ in Fig.~\ref{fig:spheres}). A representative application of this model for $M=2$ is shown in Fig.~\ref{fig:results-CS}. With the exception of the more pronounced deviation near the contact distance $r/ \sigma \approx 1$ (chiefly due to the solvation approximation, and not to the model itself), the volume-exclusion theory is again in good quantitative agreement with direct simulation results.

Since corrections at a given order to the structure furnish corrections at next order to thermodynamics, it is of interest to investigate the consequences of the above first-order structural theory to second-order corrections in the free energy, $A_2$ \cite{barker76}. For the models studied, the theory predicts $A_2$ with relative errors of the order of $10 \%$ at densities $\rho \sigma^3 \gtrsim 0.8$, the source of discrepancy at lower densities being traced back to the underlying solvation approximation (Fig.~\ref{fig:A2}), as already anticipated. It should be emphasized, however, that the volume-exclusion model above is in principle justified only at high densities, and hence the gradual breakdown of the solvation approximation at lower densities is consistent with the scope of the algebraic theory.

To this author's knowledge, the foregoing is the first structural perturbation theory of liquids to offer analytic corrections to $g_0(r)$ of such conceptual and algebraic simplicity that nonetheless yields meaningful results at high densities (compare with the superposition approximation \cite{barker76}; for algebraic corrections to thermodynamic properties at first-order, where only $g_0(r)$ is required, see the recent work of Ben-Amotz and Stell \cite{stell03}). This is achieved through the introduction of two novel ideas: the solvation approximation (Eq.~(\ref{myapprox})) -- in principle applicable to any dense liquid with either discrete or continuous potentials -- and a simple volume-exclusion model for three-particle correlations in liquids with discrete potentials (Fig.~\ref{fig:spheres}). Given its ability to easily handle wells with multiple depths, it is hoped that the methodological development put forward in this paper will help elucidate the role of length scales in liquid models with thermodynamic anomalies \cite{stanley98prl,stanley01nature}, as well as in perturbative treatments of the hydrophobic effect \cite{pratt80,hummer98}. 

\acknowledgments

The author would like to thank Attila Szabo and Gerhard Hummer for numerous discussions, and Yng-Wei Chen for a critical reading of the manuscript. This research was supported by the Intramural Research Program of the NIH, NIDDK.

\end{document}